\documentstyle[aps]{revtex}
\begin{document}

\title{An Epidemic Model on Small-World Networks and Ring Vaccination}
\author{E. Ahmed$^{1,2}$, A. S. Hegazi$^{1}$ and A. S. Elgazzar$^{3,4}$}
\address{$^{1}$Mathematics Department, Faculty of Science, Mansoura 35516,
Egypt\\
$^{2}$Mathematics Department, Faculty of Science, Al-Ain P. O. Box 17551,
U.A.E\\
$^{3}$Max Planck Institut f\"ur Physik komplexer Systeme, N\"othnitzer Str. 38,
D-01187 Dresden, Germany\\
E-mail: elgazzar@mpipks-dresden.mpg.de\\
$^{4}$Mathematics Department, Faculty of Education, El-Arish 45111, Egypt}

\maketitle

\abstract
A modified version of susceptible-infected-recovered-susceptible (SIRS) model for the outbreaks of foot-and-mouth disease
(FMD) is introduced. The
model is defined on small-world networks, and a ring vaccination programme is
included. This model can be a theoretical explanation for the nonlocal
interactions in epidemic spreading. Ring vaccination is capable of eradicating
FMD provided that the probability of infection is high enough. Also an
analytical approximation for this model is studied.

\endabstract

\section{Introduction}
Currently, there are some outbreaks of foot-and-mouth disease (FMD) in some developing and European countries. 
FMD is very infectious for cattle and sheep that are essential source for human food. The outbreaks of FMD 
cause a high loss in the national income, because export markets are closed for a country once a disease 
appears in it.

Generally animals do not have natural immunity against FMD. Also, FMD can be transmitted through contacts with 
infected animals, tools, food and aerosol. So once an infected animal is discovered, one can consider the 
whole farm is infected. There are some strategies to control an outbreak. Mass vaccination 
means vaccinating the whole population with a certain rate. But vaccination is expensive and a virus usually has 
different configurations that require different types of vaccination. Also immunity due to vaccination often 
vanishes within some months. Therefore, one has to concentrate in vaccinating animals in contact to an infected 
case. This is called ring vaccination [1,2]. It follows the following procedures:\\
\\
(1) Once an infected case is discovered, the animals at the infected farm are slaughtered.\\
\\
(2) Animals are vaccinated within a ring with a certain radius around the infected farm.\\
\\
(3) A region around the infected farm is closed off.\\

Ring vaccination always cost less than mass vaccination. Also it has been proven that ring vaccination is 
more suitable for FMD than mass vaccination [2].

Generally animals live in farms or herds forming clusters. Once a farm is infected, its nearest neighbours 
become at a high risk of infection. But this does not mean the interaction is just locally. In the 1987/1988
outbreak that occurred near Hannover, Germany [2], first two infected farms were discovered. A mass vaccination 
programme had been applied. After 3 months, another 2 infected farms were discovered at 18 Km away from the initial outbreak. 
Some days later, a third infected case was observed at a distance of 8 Km from the initial outbreak. This ensures 
the presence of nonlocal interaction in the spreading of an outbreak.

Regular graphs can display the clustering property only. On the other hand random graphs [3] can show the 
nonlocal interactions without clustering. Small-world networks (SWN) [4,5] are shown to combine both local and 
nonlocal interactions. Also the concept of SWN was applied to many different systems. The results are often 
closer to the real systems than using regular lattices [6-8].

A coupled map lattice (CML) [9] is a dynamical system with discrete time and space steps and continuous states. 
It is used to model systems that consists of coupled elements. These systems often display spatiotemporal 
chaotic behaviour. The concept of CML is used to explain some aspects in many systems in different fields like 
biology, economics, etc. [9].

There are many mathematical models were suggested to study the epidemic
spreading in a population [10-13]. According to the health of each individual, the
population is classified to susceptible (S), infected (I) or recovered (R), 
In this work, we begin by studying a simple SI model on SWN. Then a model describes the epidemic spreading of FMD 
is suggested. It is an SIRS model [10] defined on SWN and 
a ring vaccination programme is considered. Also an analytical approximated model for the epidemic spreading 
on SWN is generalized to the inhomogeneous case using some concepts of CML.

\section{Small-world networks}
The concept of SWN [4,5] is proposed to describe social networks. 
The social networks are characterized by two main properties: clustering and 
small-world effect. Small-world effect means the average distance between any 
two vertices is too short in compared with the size of the lattice. Only SWN is 
shown to satisfy the two properties together [5]. It is a connected 1-dimensional 
lattice of size $L$, 
with periodic boundary conditions. Some randomly chosen vertices are joined by 
some shortcuts to randomly selected other lattice sites. Here we consider shortcuts with length $k=1$. Let $\phi$ be the 
average number of shortcuts per bond in the lattice. Hence for large $L$, the 
probability that two random vertices are connected by a shortcut is 
$\psi \simeq 2k\phi /L$. Naturally the critical concentration of this graph $p_{c}$ is smaller than that for
the ring $p_{c}=1$, and this was derived in [11].

Motivated by a simple susceptible-infected (SI) model, the spread of an epidemic 
in a population with random susceptibility is studied on SWN. This work generalizes 
that of Moore and Newman [11]. Consider the population occupy the vertices of a SWN of size $L$. 
Each individual is assumed to have a random susceptibility $1-p(i)$, such that
$p(i)\in (0,1)$. The 
propagation rule is:\\
\\
(i) S-individual $i$ having at least one infected neighbour (shortcuting
neighbours are included, if exist) is infected if $p(i)\le q$, where $q$ is the infection rate.\\

It is relevant to estimate the number of infected persons in
this model as a function of
time [14]. Setting $q=1$ and $p(i)=0\; \forall \;i =1, 2, \ldots, L]$. Hence the number of
infected individuals will grow initially as a sphere with surface $\Gamma
_{d}t^{d-1}$, where $\Gamma_{1}=2$, $\Gamma_{2}=2\pi$, $\Gamma_{3}=4\pi$ and so on. 
This is called the primary sphere. The probability of finding a shortcut is 
$2\phi \Gamma _{d}t^{d-1}$ per unit time. Once a shortcut is reached a secondary 
sphere is formed and so on. Hence the total number of infected cases is given by
\begin{equation}
V(t)=\Gamma_{d} \int_{0}^{t} \tau ^{d-1} \biggl [1+2\phi V(t-\tau)\biggr ] d\tau.
\end{equation}
Defining $\widetilde{V}=2\phi V,\;\widetilde{t}=t[2\phi \Gamma
_{d}(d-1)!]^{1/d}$ and differentiating $d$ times with respect to $\widetilde{t}$ 
one gets
\[
\frac{\partial ^{d} \widetilde{V}}{\partial \widetilde{t}^{d}}=1+\widetilde{V}.
\]
Its solution is
\begin{equation}
\widetilde{V}(\widetilde{t})=\sum_{i=1}^{\infty} \frac{\widetilde{t}^{di}}{(di)!}.
\end{equation}

In one dimension, one has $\widetilde{V}(\widetilde{t})=\exp (\widetilde{t})-1$, 
in 2-dimensions $\widetilde{V}(\widetilde{t})=\cosh (\widetilde{t})-1$. 
For $\widetilde{t} < 1$, the number of infected individuals grow as a
power law $\widetilde{t}^{d}/d!$. While for $\widetilde{t} > 1$, it grows exponentially. 
The transition occurs at $\widetilde{t}=1$ i.e. at $t=2\phi \Gamma _{d}(d-1)!$. This result 
has an important effect in vaccination policies. It implies that vaccination should be 
administered as early as possible and with the highest possible ability to avoid reaching the 
exponential phase. Also immunizing individuals with shortcuts is more efficient than immunizing ordinary
individuals.

\section{Numerical simulations}
The susceptible-infected-recovered-susceptible (SIRS) model [10] is more closer
to FMD. The transition between the states S, I and R occur according to the following rules:\\
\\
(i) (Infection): An S-individual having at least one infected neighbour becomes
infected in the next time step with probability $q$.\\
\\
(ii) (Recovery): I-individuals are recovered by a rate $q_3$.\\
\\
(iii) (Losing immunity): The recovered individual change to the S-class by a
rate $q_2$.\\
\\
As a mean field approximation, the time rate of $S$, $I$ and $R$ are
\begin{equation}
\frac {dS}{dt}=q_2 R - qSI.
\end{equation}
\begin{equation}
\frac {dI}{dt}=qSI - q_3 I.
\end{equation}
\begin{equation}
\frac {dR}{dt}=q_3 I - q_2 R.
\end{equation}
These equations have two steady states:
\begin{equation}
S_1=N,\;\;I_1=0,\;\;R_1=0.
\end{equation}
\begin{equation}
S_2=\frac {q_3}{q},\;\;I_2=\frac {q_2}{q_2+q_3} (N-S_2),\;\;R_2=\frac {q_3}{q_2}
I_2.
\end{equation}
To get a positive value for $I_2$, $N$ must be greater than $S_2$. Then the
disease can be eradicated if $\frac {Nq}{q_3} < 1$.

This mean field approximation ignores the spatial structure of the lattice, and
is valid for the case of global interacting model. But in reality, an outbreak
spreads locally with some nonlocal interactions. To combine both local and
nonlocal interactions, we define this model on SWN. A 1-dimensional
lattice with periodic boundary conditions is assumed. A fraction of $2\phi \%$ of
its individuals is assumed to interact with a third individual (randomly chosen through
the lattice), in addition to the local interaction with the nearest neighbours.
Also a ring vaccination programme is included to control the disease propagation.
So the following procedure is added to that of the SIRS model:\\
\\
(Ring vaccination): With a probability $\sigma$, an infected case and its
nearest neighbours is changed to the R-state.\\
\\
The shortcuting neighbours are not included in this step, because they are
usually unknowns.

The simulations were carried out on a SWN with $L=5000$ and $\phi=0.05$. The
shortcuting neighbours are fixed beforehand. The values of $q_2=0.2$ and
$q_3=0.05$ are fixed through the simulations. The value of $q$ is varied $q\in
(0,1)$ and the corresponding smallest $\sigma$ sufficient to eradicate the
disease is calculated. The results are averaged over $10$ independent runs and
given in Fig. (1).

An analytical approximation for the model is given in Appendix A.
 
\section{Conclusions}
Using SWN is better than using regular or random lattices in modelling the 
outbreaks of FMD. SWN describe both local and nonlocal interactions.

For a simple SI model defined on SWN: Initially the number of infected
individuals grows as a power law, then after a critical value of time, it grows
exponentially. So vaccination should be adminstered as early as possible to
avoid reaching the exponential growth.

Ring vaccination is capable of eradicating FMD even for high infection
probability.

\appendix

\section{Analytical calculations}

Here a model is presented that approximates the epidemic spread on SWN. This
model has been used before in different contexts [12,13]. Now it is generalized 
to the inhomogeneous case. Consider $n$ patches each one contains a certain number 
of individuals (say animals). In general these patches are not identical. 
Infection spreads from infected animals within the patch and due to those 
diffusing from other patches. Then the time rate of the number of infected 
individuals in the $i$th patch is given by:
\begin{equation}
\frac {dy_{i}}{dt} = \lambda _{i}y_{i}(1-y_{i})+\mu _{i}(1-y_{i})\sum_{j\neq
i}y_{j}-\gamma _{i}y_{i}.
\end{equation}
The first term represents the infection within the patch with a rate $\lambda_i$. 
The second represents the effect of other patches both nearby and far away at a 
rate $\mu_i$. The recovery rate is represented by $\gamma_i$. Since the effect of other patches 
(second term in Eq. (A1)) is significantly smaller than the first one we expect 
that $\mu_{i} \ll \lambda _{i}$. Now the disease is eradicated if 
$y_{i}=0\;\forall i=1,2,...,n$. So the stability of this solution is studied. 
The system (A1) is a kind of CML, so we begin by presenting general stability 
results for CML. Then it is applied to Eq. (A1).

Typically a 1-dimensional CML is given by
\begin{equation}
\theta_{j}^{t+1}=(1-D)\theta_{j}^{t}+\frac{D}{2} \biggl[\theta_{j+1}^{t}+\theta
_{j-1}^{t}\biggr]+f(\theta_{j}^{t}),
\end{equation}
or
\begin{equation}
\theta _{j}^{t+1}=(1-D)f(\theta _{j}^{t})+\frac{D}{2} \biggl[f(\theta
_{j-1}^{t})+f(\theta _{j+1}^{t})\biggr],
\end{equation}
where $t=1,2,...$ and $j=1,2,...,n$. Consider the following inhomogeneous 
steady state of the system
\begin{equation}
\theta _{j}^{t}=\alpha_{j}.
\end{equation}
Linearizing around this solution, the system (A3) becomes
\begin{equation}
\varepsilon _{j}^{t+1}=(1-D)f^{\prime}(\alpha _{j})\varepsilon
_{j}^{t}+\frac{D}{2} \biggl[f^{\prime}(\alpha_{j-1})\varepsilon
_{j-1}^{t}+f^{\prime}(\alpha_{j+1})\varepsilon_{j+1}^{t}\biggr].
\end{equation}
A useful result on the eigenvalues of the system (A5) is Gerschgorin theorem
[15]. It states that the eigenvalues $\lambda$ of a square matrix 
$\left[a_{ij}\right]$ satisfy $\left| \lambda -a_{ii}\right| \leq \sum_{j\neq i}
a_{ij}$, hence
\begin{equation}
\left| \lambda \right| \leq \sum_{j}a_{ij}.
\end{equation}
Hence the steady state solution (A4) is stable if the following conditions 
are satisfied
\begin{equation}
\left| (1-D)f^{\prime}(\alpha_{j})\right| +\left| \frac {D}{2}f^{\prime}
(\alpha_{j+1})\right| +\left| \frac {D}{2}f^{\prime}(\alpha_{j-1})\right|
<1\forall j=1,2,...,n.
\end{equation}

For the homogeneous case:
\begin{equation}
\alpha_{j}=\alpha.
\end{equation}
Then the stability conditions reduce to $\left| f^{\prime}(\alpha)\right|<1$.
Condition (A7) shows how diffusion may stabilize the system, since if 
$f^{\prime}(\alpha _{j})>1$ and $f^{\prime}(\alpha_{j\pm 1})<1$. Then Eq. (A7)
may be still valid for enough large coupling.
The stability conditions for the homogeneous solution (A8) for the system
(A2) are
\begin{equation}
\left| 1-D\biggl[1-\cos (\frac{2\pi k}{n})\biggr]+f^{\prime}(\alpha)\right|<1\;\forall
k=0,1,...,n-1.
\end{equation}

Generalizing to the 2-species CML, we get
\begin{equation}
x_{j}^{t+1}=(1-D)f_{j}^{t}+ \frac {D}{2}\biggl (f_{j+1}^{t}+f_{j-1}^{t}\biggr ),
\;y_{j}^{t+1}=(1-D)g_{j}^{t}+ \frac {D}{2}\biggl (g_{j+1}^{t}+g_{j-1}^{t}\biggr ),
\end{equation}
where $f_{j}^{t}=f(x_{j}^{t},y_{j}^{t})$, $g_{j}^{t}=g(x_{j}^{t},y_{j}^{t})$.  
Consider the homogeneous steady state $x_{j}^{t}=x_{0},y_{j}^{t}=y_{0}$. 
Linearizing around it, by taking $x_{j}^{t}=x_{0}+\varepsilon_{j}^{t},\;y_{j}^{t}
=y_{0}+\eta_{j}^{t}$. Also define the doublet:
\begin{equation}
\chi _{j}^{t}=\left[ 
\begin{array}{c}
\varepsilon _{j}^{t} \\ 
\eta _{j}^{t}
\end{array}
\right].
\end{equation}
Then the linearized equations can be written in the following form
\begin{equation}
\chi_{j}^{t}=A_{j}\chi_{j}^{t}+A_{j+1}\chi_{j+1}^{t}+A_{j-1}\chi_{j-1}^{t},
\end{equation}
where
\begin{equation}
A_{k}=(1-D)\left[ 
\begin{array}{cc}
\frac {\partial f}{\partial x} & \frac {\partial f}{\partial y} \\ 
\frac {\partial g}{\partial x} & \frac {\partial g}{\partial y}
\end{array}
\right] \;\;\; {\rm at}\;(x_{0},y_{0}).
\end{equation}
This system corresponds to the direct product matrix $C=A_{j}\otimes
I+A_{j}\otimes \Sigma +A_{j}\otimes \Sigma ^{n-1}$, where
\begin{equation}
\Sigma =
\left[ 
\begin{array}{ccccc}
0 & 1 & \ldots &  & 0 \\ 
0 & 0 & 1 & \ldots & 0 \\ 
\vdots & \vdots  & \vdots &  & \vdots \\ 
&  &  &  & 1 \\ 
1 & 0 & \ldots &  & 0
\end{array}
\right].
\end{equation}
Thus the steady state of the system (A10) is stable if all the
eigenvalues $\lambda _{k}$ of the matrix $A$ satisfy 
$\left| \lambda_{k}\right| <1$, where
\begin{equation}
A=A_{j}+A_{j+1}e^{-2\pi ik/n}+A_{j-1}e^{2\pi ik/n}.
\end{equation}

The equations of a CML with delay [16] are given by
\begin{equation}
x_{j}^{t+1}=(1-D)f(x_{j}^{t})+\frac {D}{2} \left [f(y_{j-1}^{t})+
f(y_{j+1}^{t}) \right ],\;\;\;y_{j}^{t+1}=x_{j}^{t}.
\end{equation}
Repeating the above analysis, we obtain that the steady states $x_{j}^{t}
=y_{j}^{t}=x_{0}$ are stable if all the eigenvalues of the
following $2\times 2$ matrix $B=B_{j}+B_{j+1}e^{-2\pi ik/n}+B_{j-1}e^{2\pi ik/n}$
satisfy $\left| \lambda _{j,k}\right| <1$, where
\begin{equation}
B_{j}=\left[ 
\begin{array}{cc}
(1-D)f^{\prime}(x_{j}) & 0 \\ 
1 & 0
\end{array}
\right],\;\;B_{j\pm 1}=\left[ 
\begin{array}{cc}
0 & \frac{D}{2}f^{\prime}(x_{j\pm 1}) \\ 
0 & 0
\end{array}
\right].
\end{equation}

The stability conditions for the inhomogeneous solution (A4) of the
2-species system (A10) is:
\begin{equation}
\begin{array}{c}
(1-D) \left( \left| \frac {\partial f}{\partial x_{j}}\right| +\left| \frac {\partial f}{\partial y_{j}}\right|
\right)+\frac {D}{2}\left( \left| \frac {\partial f}{\partial x_{j+1}}\right| +\left| \frac {\partial
f}{\partial y_{j+1}}\right| \right) \\
+\frac {D}{2} \left( \left| \frac {\partial f}{\partial x_{j-1}}\right|
+\left| \frac {\partial f}{\partial y_{j-1}}\right| \right)<1,\;\;\forall j=1,2,...,n.
\end{array}
\end{equation}
For the delayed system (A16), the stability conditions of (A4) is:
\begin{equation}
(1-D)\left| f_{j}^{\prime}\right| +\frac {D}{2}\left| f_{j+1}^{\prime}\right|
+\frac {D}{2}\left| f_{j-1}^{\prime}\right| < 1,\;\;\; \forall j=1,2,...,n,
\end{equation}
where $f_{j}^{\prime }\equiv df/dx_{j}$.

From the above results, it is straightforward to prove the following proposition:\\
\\
\textbf {Proposition 1}: If local components are stable, then the homogeneous
steady state solution of the corresponding CML is stable.\\

Globally coupled CML are typically given by:
\begin{equation}
x_{j}^{t+1}=(1-D)f(x_{j}^{t})+\frac {D}{n-1}\sum_{k\neq j}f(x_{k}^{t}).
\end{equation}
The inhomogeneous steady state is stable if
\begin{equation}
(1-D)\left| f^{\prime}(\alpha _{j})\right| +\frac {D}{n-1}\sum_{k\neq j}\left|f^{\prime}
(\alpha _{k})\right| <1,\;\;\forall j=1,2,...,n.
\end{equation}

Continuous time CML are typically given by:
\begin{equation}
\frac {d\theta _{j}}{dt}=f(\theta _{j})+g(\theta _{j+1})+g(\theta_{j-1}).
\end{equation}
The homogeneous steady state is given by $\theta _{j}=\theta_{0} \forall j = 1,2,...,n$ and
$d\theta /dt=0$. Hence $f(\theta _{0})+g(\theta _{0})+g(\theta _{0})=0$, and it is stable if
\begin{equation}
{\rm Re}\;\left[ f^{\prime}(\theta)+2g^{\prime}(\theta)\cos \frac {2\pi r}{n}\right] < 0,\;\;r=0,1,2,...,n-1.
\end{equation}

Another form of continuous time CML is
\begin{equation}
\frac {d\theta_{j}}{dt}=f(\theta_{j})+[g(\theta_{j+1})+g(\theta_{j-1})]h(\theta_{j}).
\end{equation}
In this case the steady state is $f(\theta_{0})+[g(\theta_{0})+g(\theta
_{0})]h(\theta_{0})=0$, and it is stable if
\begin{equation}
{\rm Re}\;\left[
f^{\prime}(\theta)+2h^{\prime}(\theta)g(\theta)+2h(\theta)g^{\prime}(\theta)\cos \frac
{2\pi r}{n}\right] < 0,\;\;r=0,1,2,...,n-1.
\end{equation}

Returning to Eq. (A1), the stability of the zero solution is determined by the eigenvalues of the following matrix
\begin{equation}
A=\left[ 
\begin{array}{ccccc}
\alpha_{1} & \mu_{1} & \mu_{1} & \ldots & \mu_{1} \\ 
\mu_{2} & \alpha_{2} & \mu_{2} & \ldots & \mu_{2} \\ 
\mu_{3} & \mu_{3} & \alpha_{3} & \ldots & \mu_{3} \\ 
\vdots & \vdots & \vdots & \vdots & \vdots \\ 
\mu_{n} & \mu_{n} & \mu_{n} & \ldots & \alpha_{n}
\end{array}
\right], \; \alpha_{i}=\lambda_{i}-\gamma_{i}.
\end{equation}
For the homogeneous case where all the patches are identical, the parameters
are independent of $i$. Hence the matrix $A$ is circulant and the largest
eigenvalue is $\lambda -\gamma +\mu$. Therefore if $\lambda -\gamma <0$ and 
$\lambda -\gamma +\mu >0$, then  the disease would have been
eradicated locally, but the diffusion term may cause it to persist. This shows the
importance of the diffusion term (the long range edges in the case of SWN).

Considering the more realistic inhomogeneous case where the patches are
different. Also taking the natural assumption that $\mu_{i} \ll \lambda_{i}$,
the matrix $A$ is studied keeping only up to the second order terms in $\mu$.
Thus the Routh-Hurwitz stability conditions are that all the following
determinants are positive:
\[
\Delta_{1}=a_{1},\;\;\Delta _{2}=\left| 
\begin{array}{cc}
a_{1} & 1 \\ 
0 & a_{2}
\end{array}
\right|,\;\;\Delta _{3}=\left| 
\begin{array}{ccc}
a_{1} & 1 & 0 \\ 
a_{3} & a_{2} & a_{1} \\ 
0 & 0 & a_{3}
\end{array}
\right|, \;\;\ldots,
\]
\[
\Delta_{n}=\left| 
\begin{array}{ccccc}
a_{1} & 1 & 0 & \ldots & 0 \\ 
a_{3} & a_{2} & a_{1} & \ldots & 0 \\ 
a_{5} & a_{4} & a_{3} & \ldots & 0 \\ 
\vdots & \vdots & \vdots & \vdots & \vdots \\ 
0 & 0 & 0 & \ldots & a_{n}
\end{array}
\right|,
\]
where
\[
a_{1}=-\sum_{i}\alpha_{i},\;\;a_{2}=\sum_{j\neq i}\alpha _{i}\alpha
_{j}-\mu _{i}\mu _{j},\;\;a_{3}=-\sum_{i}\alpha _{i}\prod_{j\neq i,k\neq
i}(\alpha _{k}\alpha _{j}-\mu _{k}\mu _{j}),
\]
\[
a_{4}=\sum_{i,l}\alpha _{i}\alpha _{l}\prod_{j\notin \{i,l\},k\notin
\{i,l\}}(\alpha _{k}\alpha _{j}-\mu _{k}\mu _{j}),\;\; \ldots,
\]
\[
a_{n}=(-1)^{n}\left [\prod_{i}\alpha _{i}-\sum_{i,l}\mu _{i}\mu
_{l}\prod_{j\notin \{i,l\},k\notin \{i,l\}}\alpha _{k}\alpha
_{j}\right].
\]
The effect of ring vaccination is expected to reduce $\lambda _{i}$ not $
\mu _{i}$.

\newpage
\begin{center}
{\bf Figure Caption}        
\end{center}
Fig. 1: The relation between the infection probability $q$ and the smallest probability
$\sigma$ sufficient to eradicate a disease.


\begin{thebibliography}{99}
\bibitem[1]{1}  D. Greenhalgh, Comm. Stat. Stoch. Models 2, 339 (1986).

\bibitem[2]{2}  J. M\"uller , B. Sch\"ofisch and M. Kirkilionis, J. Math. Biol 41, 143 (2000).

\bibitem[3]{3}  B. Bollobas, Random Graphs (Academic Press, New York, 1985).

\bibitem[4]{4}  D. J. Watts and S. H. Strogatz, Nature 393, 440 (1998).

\bibitem[5]{5}  M. E. J. Newman  and D. J. Watts, Phys. Lett. A 263, 341 (1999).

\bibitem[6]{6}  E. Ahmed and H. A. Abdusalam, Eur. Phys. J. B 16, 569 (2000).

\bibitem[7]{7}  N. Mathias and V. Gopal, Phys. Rev. E 63, 21117 (2001).

\bibitem[8]{8}  A. S. Elgazzar, Physica A (submitted, 2001).

\bibitem[9]{9}  K. Kaneko, "Theory and Applications of Coupled Map lattices", (J. Wiley \&
Sons, Chichester, 1993).

\bibitem[10]{10}  L. Edelstein-Keshet, "Mathematical Models in Biology", (Random House,
New York, 1988).

\bibitem[11]{11}  C. Moore and M. E. J. Newman, Phys. Rev. E 61, 5678 (2000).

\bibitem[12]{12}  A. Lajmanovich and J. A. Yorke, Math. Biosci. 28, 221 (1976).

\bibitem[13]{13}  F. Ball, Math. Biosci. 156, 41 (1999).

\bibitem[14]{14}  C. F. Moukarzel, Phys. Rev. E 60, 6263 (1999).

\bibitem[15]{15}  N. Chatterjee and N. Gupte, Physica A 224, 422 (1996), and nlin/9702015.

\bibitem[16]{16}  D. H. Zanette "Structures and Propagation in Globally Coupled Systems
with Time Delayes", cond-mat/0003174 (2000). 


\end{thebibliography}
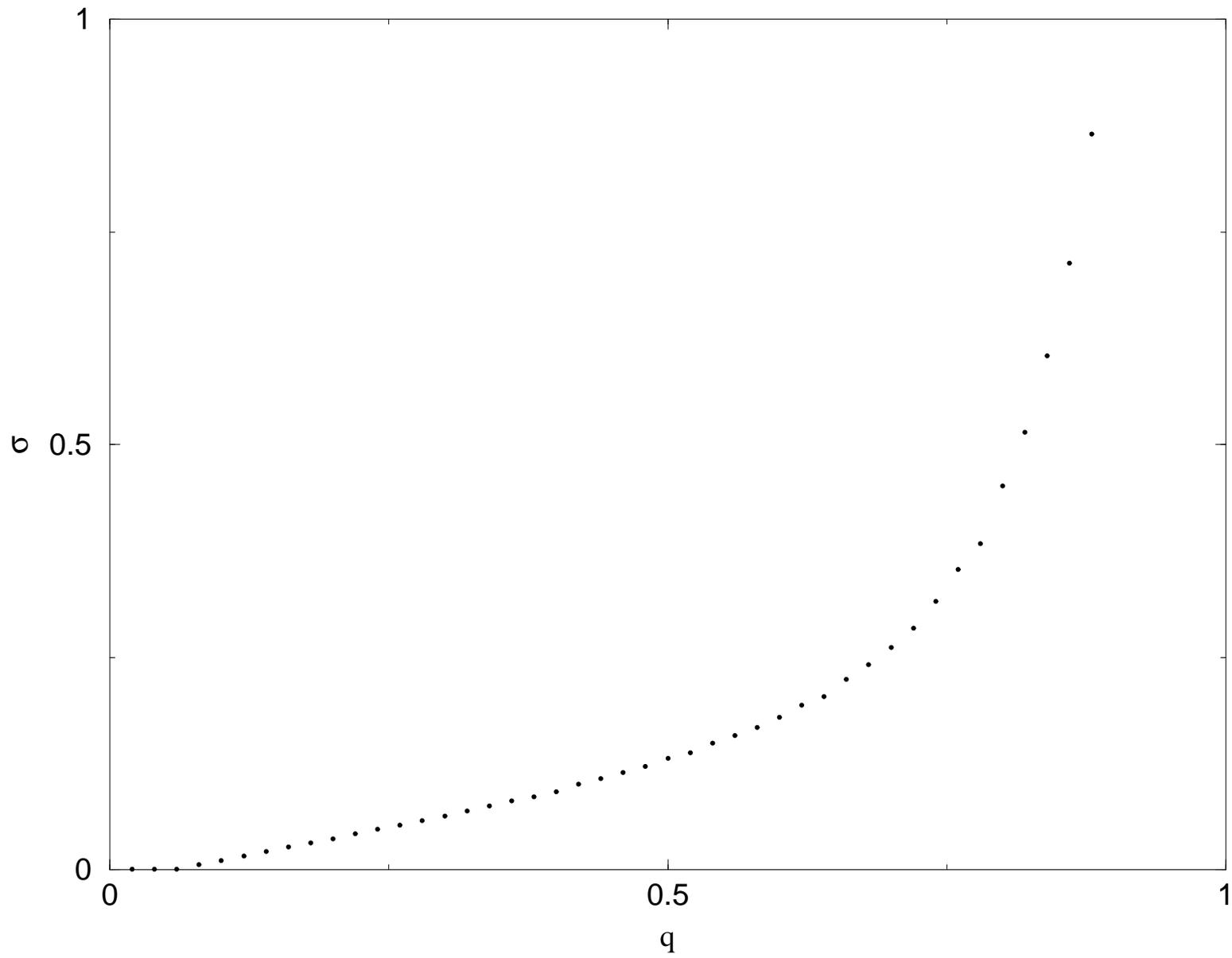
\end{document}